\documentclass[twoside,12pt]{article}
\usepackage{CJK}
\usepackage{indentfirst}
\usepackage{bm}
\usepackage{graphicx}
\usepackage{amsmath}
\usepackage{latexsym}

\begin{document}
%\begin{CJK*}{GBK}{song}

\thispagestyle{empty} \vspace*{0.1cm}\hbox
to\textwidth{\vbox{\hfill\sf Communications in Theoretical Physics, Vol. 59, No. 1, January 15, 2013, 22¨C26\hfill}}
\par\noindent\rule[3mm]{\textwidth}{0.2pt}\hspace*{-\textwidth}\noindent
\rule[2.5mm]{\textwidth}{0.2pt}

\begin{center}
On models of nonlinear evolution paths in adiabatic quantum algorithms$^{*}$
\end{center}

\footnotetext{\hspace*{-.45cm}\footnotesize $^*$Project supported by the National Natural Science Foundation of China (Grant No. 61173050).}
\footnotetext{\hspace*{-.45cm}\footnotesize $^\dagger$Corresponding author: lusongfeng@hotmail.com }

\begin{center}
\rm SUN Jie $^{\rm a)}$, \ \ LU Song-Feng $^{\rm a)}$$^{\dagger}$,  \ Samuel L. Braunstein$^{\rm b)}$
\end{center}

\begin{center}
\begin{footnotesize} \sl
$^{\rm a)}$${School \ of\ Computer\ Science, Huazhong\ University\ of\ Science\ and\ Technology,\ Wuhan \ 430074,\ China}$\\
$^{\rm b)}$${Department \ of\ Computer\ Science, University\ of\ York,\ York \ YO10 \ 5DD,\ United \ Kingdom}$ \\

\end{footnotesize}
\end{center}

\begin{center}
\footnotesize (Received June 25, 2012; revised manuscript received November 9, 2012)
\end{center}

\vspace*{2mm}

\begin{center}
\parindent 20pt\footnotesize
\begin{abstract}

In this paper, we study two different nonlinear interpolating paths in adiabatic evolution algorithms for solving a particular class of quantum search problems where both the initial and final Hamiltonian
are one-dimensional projector Hamiltonians on the corresponding ground state. If the overlap between the initial state and final state of the quantum system is not equal to zero, both of these models can provide a constant time speedup over the usual adiabatic algorithms by increasing some another corresponding ``complexity''. But when the
initial state has a zero overlap with the solution state in the problem, the second model leads to an infinite time complexity of the algorithm for whatever interpolating functions being applied while the first one can still provide a constant running time. However, inspired by a related reference, a variant of the first model can be constructed which also fails for the problem when the overlap is exactly equal to zero if we want to make up the ``intrinsic" fault of the second model---an increase in energy. Two concrete theorems are given to serve as explanations why neither of these two models can improve the usual adiabatic evolution algorithms for the phenomenon above. These just tell us what should be noted when using certain nonlinear evolution paths in adiabatic quantum algorithms for some special kind of problems.

\end{abstract}
\end{center}

\begin{center}
\begin{minipage}{15.5cm}

\end{minipage}
\begin{minipage}[t]{2.3cm}{\bf Key words:}\end{minipage}
\begin{minipage}[t]{13.1cm}
adiabatic evolution, \ nonlinear evolution paths, \ quantum computing.
\end{minipage}\par\vglue8pt
\end{center}

\section{Introduction}
Quantum adiabatic evolution \cite{Farhi2001} has been developed as an alternative to the traditional circuit model of quantum computation and the equivalence between these two models has been proven recently \cite{Mizel2007}, \cite{Aharonov2007}. But adiabatic quantum computation has demonstrated some of its own advantages, especially like the robustness against different kinds of perturbations, such as decoherence \cite{Johan2005}, and unitary control errors \cite{Childs2001}.

The main idea behind an adiabatic quantum algorithm can be summarized as follows: prepare a quantum system in the ground state of a simple initial Hamiltonian, then slowly change this Hamiltonian to a final Hamiltonian whose ground state encodes the solution to the problem that will be solved. More precisely, assume the Hamiltonian
$H(t)$ of a quantum system can fully characterize the time evolution of the system. Let $|\varphi(t)\rangle$ be the state of the system which is evolving according to the Schr\"odinger equation
\begin{equation}
i\frac{d}{dt}|\varphi(t)\rangle=H(t)|\varphi(t)\rangle.
\end{equation}
Define $|E(0,t)\rangle$ and $|E(1,t)\rangle$ as the ground state and first excited state of the system Hamiltonian $H(t)$ with corresponding eigenvalues $E(0,t)$ and $E(1,t)$, respectively. The quantum adiabatic theorem \cite{Messiah1999} tells that if we prepare the system in the ground state of the initial Hamiltonian $H_{i}$
and let it evolve slowly along some path to the final Hamiltonian $H_{f}$, then
\begin{equation}
|\langle E(0,T)|\varphi(T)\rangle|^{2}\geq1-\varepsilon^{2}(0<\varepsilon\ll1),
\end{equation} provided that
\begin{equation}
\frac{\Delta_{max}}{\delta_{min}^{2}}\leq\varepsilon,
\end{equation}in which
\begin{equation}
\Delta_{max}=\underset{0\leq t\leq T}\max\Bigg|
\langle E(1,t)\big|\frac{dH(t)}{dt}\big|E(0,t)\rangle\Bigg|
\end{equation}
is a measurement of the evolving rate of the Hamiltonian, and
\begin{equation}
\delta_{min}=\underset{0\leq t\leq T}\min[E(1,t)-E(0,t)]
\end{equation}
is the minimum gap between the two lowest eigenvalues. In general, the required run time $T$ for a typical adiabatic algorithm will be mainly determined by $\delta_{min}$ so long as $\Delta_{max}$ is polynomially bounded. For the sake of convenience, the time-dependent Hamiltonian $H(t)$ can be reparametrized as
\begin{equation}
\widetilde{H}(s)=H(t/T), \ for \ 0\leq s \leq 1.
\end{equation}
Usually, the instantaneous Hamiltonian $\widetilde{H}(s)$ which connects the initial Hamiltonian $H_{i}$ and final Hamiltonian $H_{f}$ of the quantum system is a linear adiabatic evolution path
\begin{equation}
\widetilde{H}(s)=(1-s)H_{i}+sH_{f}.
\end{equation}
But this simple form for the interpolating Hamiltonian may sometimes have some limitations for some special problems, as we will show.

In fact, there is no reason not to try paths involving terms that are not linear combinations of the initial and final Hamiltonians. The adiabatic algorithm will work taking any path $\widetilde{H}(s)$, as long as the adiabaticity condition is satisfied: the ground state at $s=0$ is the initial state and the ground state at $s =1$ is the solution state. In Ref. \cite{Farhi2002}, the authors had proposed the following general form of time-dependent Hamiltonian,
\begin{equation}
\label{addingdrivingHamiltonian}
\widehat{H}(s)=(1-s)H_{i}+sH_{f}+s(1-s)H_{e},
\end{equation}
in which $H_{e}$ is an extra piece of the Hamiltonian that is turned off at the beginning and end of the adiabatic evolution. It was found this nonlinear evolution can change the performance of the algorithm from unsuccessful to successful when solving some special problems, such as those discussed in Ref. \cite{Farhi2002Goldstone}. As a result, when applying this method to the quantum search problem \cite{Grover1997}, by a suitable choice of the time-dependent Hamiltonian, it is possible to do the calculation in constant time, independent of the number of items in the database. However, in this case, the original time-complexity is replaced by the complexity of implementing the driving Hamiltonian $H_{e}$ \cite{Andrecut2004}.

In Ref. \cite {Das2003}, a general class of models for the time-dependent Hamiltonian was proposed, which has a form like:
\begin{equation}
\label{gerenalmodelHamiltonian}
\overline{H}(s)=f(s)H_{i}+g(s)H_{f},
\end{equation}
where $f(s)$ and $g(s)$ are arbitrary functions of time, subject to the boundary conditions
\begin{equation}
\label{boundarycondition}
f(0)=g(1)=1, \qquad f(1)=g(0)=0.
\end{equation}
When performing unstructured adiabatic quantum search of a database with $N$ items, it was found that by increasing the lowest eigenvalue of the time-dependent Hamiltonian temporarily to a maximum of $\propto\sqrt{N}$, it was possible to do the calculation in constant time as well.

In this paper, we study a special class of quantum search problems where both the initial and final Hamiltonian are one-dimensional projector Hamiltonians on the corresponding ground state by using the above two different approaches. When the initial state has a nonzero overlap with the final state, both of these approaches can
provide a constant running time by increasing some another corresponding ``complexity''. However, if the overlap is exactly equal to zero, we find that only the adding a driving Hamiltonian approach whose form is given by Eq.~(\ref{addingdrivingHamiltonian}) can still work while the other approach fails no matter what the
interpolating functions chosen for $f(s)$ and $g(s)$. But at the same time, if we want to eliminate the pitfall of the general class of models for the time-dependent Hamiltonian---an increase in energy for the system, we find that a variant of the driving Hamiltonian model can be built which also lead to an infinite time complexity for the problem when the initial state of the system is orthogonal to the solution state. These two facts just tell what we should note when using some nonlinear interpolating paths in adiabatic evolution algorithms for some special kind of problems.

The organization of the current paper is as follows: in part 2, we define the problem that will be studied throughout the paper. The two different approaches for the algorithmic performance speedup of this problem as compared to a local adiabatic \cite{Roland2002} or partial adiabatic evolution algorithm \cite{Tulsi2009} are discussed detailedly in part 3. Two concrete theorems are given to show why both of the variant version of the first model and the second model fails for the problem when the initial state is orthogonal to the final state in part 4. We summarize and draw some conclusions in the last part.

\section{The problem}
For the purpose of illustrating our main idea, we discuss the adiabatic evolution for the problem studied in the current context which has a form for the initial and final Hamiltonian as
\begin{equation}
H_{i}=I-|\alpha\rangle\langle\alpha|,\qquad
H_{f}=I-|\beta\rangle\langle\beta|.
\end{equation}
It is easy to know that in fact our this defined problem can be seen as a kind of abstraction of quantum search problem. Now the problem is how long it will take us to evolve the initial state $|\alpha\rangle$ to the final state $|\beta\rangle$ by applying a nonlinear global adiabatic evolution algorithm. In fact, if a simple linear adiabatic algorithm \cite{Farhi2000} is applied, it is easy to know the running time can be estimated as
\begin{equation}
T_{1}=O(|a|^{-2}),
\end{equation}where
\begin{equation}
a=\langle\alpha|\beta\rangle.
\end{equation}
While a local adiabatic algorithm \cite{Roland2002} or a partial adiabatic algorithm \cite{Tulsi2009} is used for the same problem, it is not difficult to see that the time complexity is
\begin{equation}
T_{2}=O(|a|^{-1}),
\end{equation}
which provides a quadratic speedup over the global adiabatic evolution. But can we further reduce the time consumption for the problem? Inspired by the two different approaches introduced in the first section, we give the answer to this question in the positive side, as we will show in detail below.
\section{Two different models for the algorithmic speedup of the problem}
In this section, we show how the two different strategies introduced above can provide a speedup over the usual adiabatic algorithm such as global adiabatic, local adiabatic or partial adiabatic algorithms in subsection 1 and 2, respectively. Before we start, we introduce the following notations of a reduced two dimensional-Hilbert space to simplify the calculation of the two lowest eigenvalues of the Hamiltonians that will be shown:
\begin{equation}
|1\rangle=|\alpha\rangle, \ |2\rangle=\frac{1}{b}(|\beta\rangle-a|\alpha\rangle),
\end{equation}
in which
\begin{equation}
b=\sqrt{1-|a|^{2}}.
\end{equation}
Thereby, we can get
\begin{equation}
|\alpha\rangle=|1\rangle, \ |\beta\rangle=a|1\rangle+b|2\rangle.
\end{equation}
\subsection{Adding a driving Hamiltonian}
Consider a quantum Hamiltonian which has the form of Eq.~(\ref{addingdrivingHamiltonian}) for solving the problem,
in which $H_{e}$ is specified as:
\begin{equation}
H_{e}=|\alpha\rangle\langle\beta|+|\beta\rangle\langle\alpha|.
\end{equation}
Now the Hamiltonian $\widehat{H}(s)$ in Eq.~(\ref{addingdrivingHamiltonian}) has a matrix form
\begin{equation}
\widehat{H}(s)=\left(\begin{array}{c c}
sb^{2}+2s(1-s)Re(a) \ & sb(1-s-a) \\
sb(1-s-a^{*})  \ & 1-sb^{2}
\end{array}\right),
\end{equation}by the notations being introduced at the beginning of this section,
and the eigenvalues of it are easy to obtain:
\begin{equation}
E_{0}(s)=\frac{A-B}{2}, \ E_{1}(s)=\frac{A+B}{2},
\end{equation}in which we have used the following marks for simplicity:
\begin{eqnarray}
A=1+2s(1-s)Re(a),\\
B=\sqrt{1-4s(1-s)\Big\{Re(a)+b^2 +s(1-s)[Im(a)]^{2}-s(1-s)}\Big\}.
\end{eqnarray}
We therefore can get the energy gap of the system Hamiltonian:
\begin{equation}
E_{1}(s)-E_{0}(s)=\sqrt{1-4s(1-s)\Big\{Re(a)+b^2 +s(1-s)[Im(a)]^{2}-s(1-s)}\Big\}.
\end{equation}
And the minimum energy gap reads
\begin{equation}
\delta_{min}=\underset{0\leq s\leq1}
\min[E_{1}(s)-E_{0}(s)]=\sqrt{|a|^{2}-Re(a)-\frac{1}{4}[Im(a)]^{2}+\frac{1}{4}}.
\end{equation}
Thereby, when the overlap $a$ is tending to zero or exactly equal to it, the time complexity for the algorithm tends to a constant. Simultaneously, it is obvious that the usual adiabatic algorithms such as linear adiabatic evolution etc. have a bad performance at this time. In fact, if $a=0$ holds, this will lead to the total failure of all
the usual adiabatic algorithms---i.e., infinite time complexity, which can be seen as follows:
\begin{equation}
\widetilde{H}(s)=
\left(\begin{array}{c c}
sb^{2} \ & -sab \\
-sa^{*}b \ & 1-sb^{2}
\end{array}\right),
\end{equation}
so the energy gap of it now reads:
\begin{equation}
\delta_{min}'=\underset{0\leq s\leq1}\min\sqrt{1-4s(1-s)(1-|a|^{2})}=0.
\end{equation}
But in the meanwhile, an obvious fact should be easy to observe: the added driving Hamiltonian in this model also takes additional ``complexity'' to implement compared with the widely studied adiabatic evolution procedure in the literature. However, this just implies more feasibilities can be achieved for the adiabatic evolution based quantum algorithms than its equivalent quantum
algorithms based on the circuit model.

\subsection{Using a more general model of interplaiting path}
Now we turn to another approach to solve the problem. Consider the general interpolating form for the Hamiltonian $\overline{H}(s)$ in Eq.~(\ref{gerenalmodelHamiltonian}), which has a matrix form in the reduced two-dimensional Hilbert space defined before:
\begin{equation}
\overline{H}(s)=\left(\begin{array}{c c}
gb^{2} \ & -gab \\
-ga^{*}b  \ & f+g|a|^{2}
\end{array}\right),
\end{equation}
and its two corresponding eigenvalues are therefore given by
\begin{eqnarray}
\label{energygapI}
\overline{E}_{0}(s)&=&\frac{(f+g)-\sqrt{(f-g)^{2}+4fg|a|^{2}}}{2}, \\
\label{energygapII}
\overline{E}_{1}(s)&=&\frac{(f+g)+\sqrt{(f-g)^{2}+4fg|a|^{2}}}{2}.
\end{eqnarray}So, we have
\begin{equation}
\delta_{min}=\underset{0\leq s\leq1}
\min[\overline{E}_{1}(s)-\overline{E}_{0}(s)]=|a|+\frac{x}{2}|a|,
\end{equation}
when we set
\begin{equation}
\label{interpolatingparameters}
f(s)=1-s+xs(1-s),\qquad g(s)=s+xs(1-s).
\end{equation}
It is easy to see that if we choose $x=\frac{1}{|a|}$, the running time of the algorithm tends to a constant when the overlap $a$ tends to zero. But simultaneously, it can be found that the ground state energy $\overline{E}_{0}(s)$ grows with a maximum value proportional to $\frac{1}{|a|}$ at the point $s=\frac{1}{2}$ as well before it returns
to zero at the end of the adiabatic evolution. In contrast, when the free parameter ``$x$'' disappears, this model has degenerated to the linear adiabatic evolution version, and the performance of it is very bad at this time(the overlap $a$ tends to zero) as can be seen from the preceding subsection, although the ground state energy of the quantum system tends to a constant at the minimum time point. In fact, when the parameters ``f(s)" \&``g(s)" are specified as in Eq.~(\ref{interpolatingparameters}), we have the following even more accurate relation between this general model of interplaiting path and the usual linear adiabatic evolution: if the minimum gap between the ground state and the first excited state of the former one increases to an amount approximate to ``x" larger than the second one, the corresponding ground state energy is enlarged to the same amount accordingly. And this is consistent with our intuition: adiabatic evolution performs slowly around the transition point where the energy gap is small, and if large energy is injected into the quantum system, it will be helpful for improving the performance of the adiabatic algorithm.

What about the time complexity of this general class of model of adiabatic evolution algorithm for the current problem when the overlap $a$ is exactly equal to zero? Obviously, Eq.~(\ref{interpolatingparameters}) will still give an infinite time complexity as the usual adiabatic algorithms do no matter what value is taken for the free parameter ``$x$''. But what about the result if other functions for $f(s)$ and $g(s)$ are chosen? We will discuss this in the next section.

\section{The theorems}
In fact, we can prove the second approach always fails for the problem when the initial state has a zero overlap with the final state.

\textbf{\emph{Theorem 1}} The second model of the interpolating Hamiltonian discussed in section 3.2 will always cause an failure the adiabatic algorithm for whatever functions of $f(s)$ and $g(s)$ are selected when it is used to solve the problem above if the initial state is orthogonal to the final solution state.

\textbf{\emph{Proof}} By Eq.~(\ref{energygapI}) and Eq.~(\ref{energygapII}), it is known that once the function $f(s)$ and $g(s)$ are chosen, the minimum energy gap between the lowest and second lowest eigenvalues is
\begin{equation}
\delta_{min}=\underset{0\leq s\leq1}
\min[\overline{E}_{1}(s)-\overline{E}_{0}(s)]
=\underset{0\leq s\leq1}\min|f(s)-g(s)|.
\end{equation}
From the boundary conditions given in Eq.~(\ref{boundarycondition}), a new function can be constructed as
\begin{equation}
F(s)=f(s)-g(s),
\end{equation}
and it follows that
\begin{equation}
F(0)=f(0)-g(0)=1, \ F(1)=f(1)-g(1)=-1.
\end{equation}
As functions $f(s)$ and $g(s)$ are both continuous and by ``Intermediate Value Theorem'' in mathematics, we immediately know that there must exist a point for which
\begin{equation}
F(s)=0,
\end{equation}
holds for some $0<s<1$. This just implies an infinite time complexity for the adiabatic algorithm and the proof is completed . \hfill$\Box$

As we have already seen from section 3.2, the speed up of the adiabatic algorithm for the problem requires an
increase in energy, at least temporarily. If we want to keep the ground state energy zero as stated in Ref. \cite{Das2003}, we could use the following adiabatic Hamiltonian as shown in the next theorem, which can be seen as a variant of the first model discussed in section 3.1. However, it also has the limitation that it fails for the problem when the overlap between the initial state and final state is exactly equal to zero.

\textbf{\emph{Theorem 2}} The following adiabatic Hamiltonian form still fails for the problem defined above when the initial state has a zero overlap with the final state:
\begin{equation}
\widehat{H'}(s)=f(s)H_{i}+g(s)H_{f}-h(s)I,
\end{equation}
in which $f(s)$ and $g(s)$ are arbitrary continuous functions, and $h(s)$ is given by
\begin{equation}
h(s)=\frac{f(s)+g(s)-|f(s)-g(s)|}{2}.
\end{equation}

\textbf{\emph{Proof}} Firstly, it can be easily verified that $h(s)$ satisfies the correct boundary condition when adding an extra Hamiltonian model, the driving item only appears during the adiabatic evolution process:
\begin{equation}
h(0)=h(1)=0.
\end{equation}
By the same method as before, we can write out the matrix form for the Hamiltonian $\widehat{H'}(s)$ as
\begin{equation}
\widehat{H'}(s)=\left(\begin{array}{c c}
g-h(s) \ & 0 \\
0  \ & f-h(s)
\end{array}\right).
\end{equation}
It follows that the minimum energy gap can also be easily obtained yielding
\begin{equation}
\delta_{min}=\underset{0\leq s\leq1}\min|f(s)-g(s)|.
\end{equation}
By the same proof progress of theorem 1, we obtain the conclusion that the time complexity for this adiabatic algorithm is also infinite. \hfill$\Box$\\\\

\section{Conclusion}
We have studied two different models of nonlinear paths in quantum adiabatic evolution for solving a special kind of quantum search problem in which both the initial and final Hamiltonian are a one-dimensional projector Hamiltonian on the corresponding ground state. When the initial state has a nonzero overlap with the solution
state, both of these approaches can provide a constant running time although the mechanisms behind them are different: in the first approach, the time complexity has been replaced by the ``complexity'' of implementing the driving Hamiltonian that is added while the ground state energy of the quantum system grows maximally proportional
to a corresponding quantity temporarily before returning to zero at the end of the adiabatic evolution in the second model. But when the overlap between the initial state and the final state is exactly equal to zero, only the first approach can still work for the problem. A concrete theorem is proven to explain why the second approach fails for whatever
interpolating functions are selected in that model. However, we have also shown that a variant version of the adding a driving Hamiltonian model which was mentioned in some related literature, will still lead to an infinite time complexity for the problem in the occasion that the overlap is equal to zero when we want to avoid an increase in energy temporarily. Using some nonlinear evolution paths in adiabatic algorithms may not always be helpful for certain special kind of problems, and we just should pay attention to this phenomenon when designing the adiabatic quantum algorithms.

%\end{CJK*}
\end{document}